\documentclass{article}
\usepackage{ijcai20}

\usepackage{times}

\usepackage{soul}
\usepackage{url}
\usepackage[hidelinks]{hyperref}
\usepackage[utf8]{inputenc}
\usepackage[small]{caption}
\usepackage{graphicx}
\usepackage{amsmath}
\usepackage{booktabs}
\usepackage{multirow}
\usepackage{multicol}
\usepackage{subfigure}
\usepackage{tikz}
\usepackage{calc}
\usepackage{bbding}
\usepackage{pifont}
\usepackage{wasysym}
\usepackage{amssymb}
\newtheorem{definition}{Definition}
\usepackage{bibspacing}
\usepackage{fancyhdr}
\usepackage{hyperref}

\title{Deep Learning for Community Detection: 
Progress, Challenges and Opportunities}

\author{
Fanzhen Liu$^1$\and
Shan Xue$^{1,2}$\and
Jia Wu$^1$\footnote{Corresponding Author}\and
Chuan Zhou$^3$\and
Wenbin Hu$^4$\and \\
Cecile Paris$^{2,1}$\and
Surya Nepal$^{2,1}$\and
Jian Yang$^1$\and
Philip S. Yu$^5$\\
\affiliations
$^1$ Department of Computing, Macquarie University, Sydney,  Australia\\
$^2$ CSIRO's Data61, Sydney, Australia\\
$^3$ Academy of Mathematics and Systems Science, Chinese Academy of Sciences, Beijing, China\\
$^4$ School of Computer Science, Wuhan University, Wuhan, China \\
$^5$ Department of Computer Science, University of Illinois at Chicago, Chicago, USA \\
\emails
\{fanzhen.liu@hdr., jia.wu@, jian.yang@\}mq.edu.au, zhouchuan@amss.ac.cn, hwb@whu.edu.cn, \{emma.xue, surya.nepal, cecile.paris\}@data61.csiro.au, psyu@uic.edu
}

\begin{document}

\maketitle

\thispagestyle{fancy}       
\fancyhead{}
\lhead{This paper was published by IJCAI-2020 with doi:\href{https://www.ijcai.org/Proceedings/2020/0693.pdf}{10.24963/ijcai.2020/693}}
\fancyfoot{}
\lfoot{The published version of this paper is available at \href{https://www.ijcai.org/Proceedings/2020/0693.pdf}{https://www.ijcai.org/Proceedings/2020/0693.pdf}}

\renewcommand{\headrulewidth}{0.5pt}
\renewcommand{\footrulewidth}{0.5pt}

\pagestyle{empty} 

\begin{abstract}
As communities represent similar opinions, similar functions, similar purposes, etc., community detection is an important and extremely useful tool in both scientific inquiry and data analytics. However, the classic methods of community detection, such as spectral clustering and statistical inference, are falling by the wayside as deep learning techniques demonstrate an increasing capacity to handle high-dimensional graph data with impressive performance. Thus, a survey of current progress in community detection through deep learning is timely. Structured into three broad research streams in this domain – deep neural networks, deep graph embedding, and graph neural networks, this article summarizes the contributions of the various frameworks, models, and algorithms in each stream along with the current challenges that remain unsolved and the future research opportunities yet to be explored.
\end{abstract}

\section{Introduction}
The two basic elements of a network/graph are nodes and edges. From the perspective of connectedness and density, communities are known as locally dense connected subgraphs or clusters of nodes \cite{FORTUNATO20161}. In addition to the internal cohesion of subgraphs, their separation from each other should also be taken into account. To this end, graph theory sets out two specific rules for determining which of these nodes and edges to form a community: 1) nodes in a community are densely connected; and 2) nodes in different communities are sparsely connected. An easy and popular understanding is that communities are subgraphs holding more internal connections than external connections.

\begin{figure}[!t]
    \centering
    \includegraphics[width=0.48\textwidth]{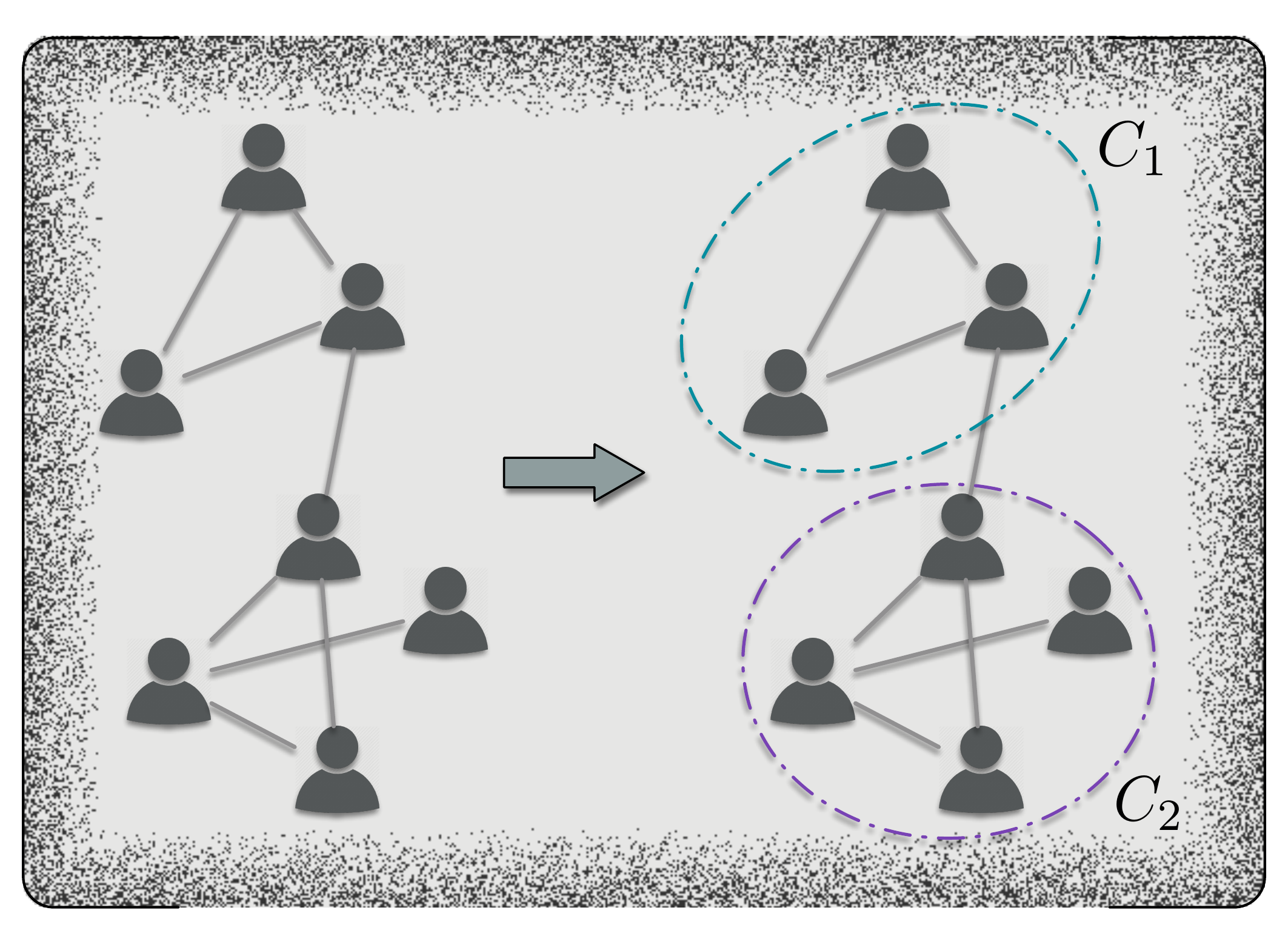}
    \caption{An example of community detection in a social network. The network is divided into two communities according to the closeness of the individuals: community $C_1$ includes three nodes, and community $C_2$ includes four nodes.}
    \label{community_detection}
\end{figure}

In the real world, the nodes in a community can share common properties or serve similar functions \cite{fortunato2010community}, and finding these commonalities is the strategic key to almost every community detection strategy that exists today. Community detection, or more specifically, clustering nodes based on a similar feature or set of features, helps us understand the inherent patterns and functions of networks. For example, community detection in protein-protein interaction (PPI) networks reveals proteins with similar biological functions \cite{Chen2006Bio}. In citation networks, community detection determines the importance, interconnectedness, and evolution of research topics \cite{CHEN2010278}. In enterprise networks, employees can be grouped into communities by studying company's offline internal sources and online enterprise social relationships \cite{7929975}. In social networks like Twitter and Facebook, users with common interests or mutual friends may be members of the same community \cite{Yang6729613}, as shown in Figure \ref{community_detection}.

Most traditional methods of community detection are based on statistical inference and conventional machine learning. For example, one of the most widely-used methods to detect communities and describe how they are formed is stochastic block model \cite{PhysRevE.83.016107}. However, despite its past popularity, this strategy struggles in the face of today’s complex datasets and complicated social scenarios. In conventional machine learning, detecting communities has generally been conceived as a clustering problem on graphs. But these approaches are highly dependent on the characteristics of the data. For example, spectral clustering \cite{ng2002spectral}, which uses eigenvectors to partition nodes into communities, does not perform well with sparse networks. 

It is becoming ever-clearer that the increasing scale of networks and dimensionality of data demand more powerful techniques to maintain effective and efficient performance with feasible computation speeds \cite{Rosvall1118}.

For now at least, deep learning \cite{DeepLearning} is the solution. With deep learning, computational models can learn representations of data at multiple levels of abstraction, which is perfectly suited to network data. Moreover, its ability to learn nonlinear features is greatly advancing, and it has been successful in a wide range of fields where the data have internal relationships, such as computer vision and natural language processing. Further, multi-layer deep neural networks can reduce the dimensionality of the data \cite{ZHANG201829}, which widens the potential scope of network analysis tasks like community detection, node classification, and link prediction.

In this survey, we focus on new trends in deep learning for community detection. Our investigation is broken into three main parts: 

\begin{enumerate}
\item [1)] We review and assess the advantages of the various deep learning methods for community detection;
\item [2)] We summarize and categorize the sate-of-the-art studies from a technical perspective; and 
\item [3)] We identify and discuss the technical challenges yet to be solved, along with suggestions of promising opportunities for future works. 
\end{enumerate}

To the best of our knowledge, this is the first work to provide a comprehensive review of deep learning for community detection. Most of the papers we surveyed were published in recent influential international conferences in the domains of artificial intelligence, machine learning, data mining, and web services, \textit{e.g.,} NIPS, ICLR, AAAI, IJCAI, KDD, ICDM, CIKM, ICDE, and WWW. Articles in high-quality peer-reviewed journals were also included. 

The information in this survey should directly support researchers and technology experts to understand the past, current, and future trends in fields of deep learning for community detection.

The remainder of this survey is organized as follows. Section \ref{Sec2} clarifies the concepts of community detection. Section \ref{Sec3} explains the reasons why today’s community detection methods must be based on deep learning. Novel deep learning-based community detection methods are surveyed in Section \ref{Sec4}. Section \ref{Sec5} discusses the remaining challenges in this field given recent progress, along with suggestions for future research opportunities for academics and domain experts. Finally, Section \ref{Sec6} concludes the survey.

\section{Community Detection}
\label{Sec2}
In this paper, a network is a special graph that abstracts complex relationships in real-world systems, such as the internets, academic collaborations, or social groups. 

\begin{definition}
\textbf{(Network)}
Following graph theory, a weighted network is represented as $G = (V,E,W)$ and an unweighted network is represented as $G=(V,E)$, where $V$ and $E$ denote the set of nodes and edges, respectively, and $W$ denotes the corresponding weights of $E$ in terms of the intensity or capacity of the connections. In an unweighted network, $W$ is regarded as 1 and can be removed from $G$.
\end{definition}

A subgraph $g \subseteq G$ is a partition of a graph that retains the original network structure. The division of subgraphs follows pre-defined rules, and different rules may result in different forms of subgraph. A community is a type of subgraph that represents a real social phenomenon. In other words, a community is a group of people or objects that share common characteristics. 

\begin{definition}
\textbf{(Community)}
Communities are the subgraphs in a network where the nodes share dense connections. Sparsely-connected nodes delineate communities. Here, we use $\mathcal{C} = \left\{C_1, C_2, ..., C_k \right\}$ to denote a set of $k$ communities divided from a network $G$, where ${C}_i$ is the $i$th community from this network partition. A node $v$ clustered into the community $C_i$ satisfies the condition that the internal degree of each node inside the community exceeds its external degree.
\end{definition}

Thus, the goal of community detection is to discover the communities  $\mathcal{C}$ in a network $G$.

\begin{figure*}
    \centering
    \includegraphics[width=1\textwidth]{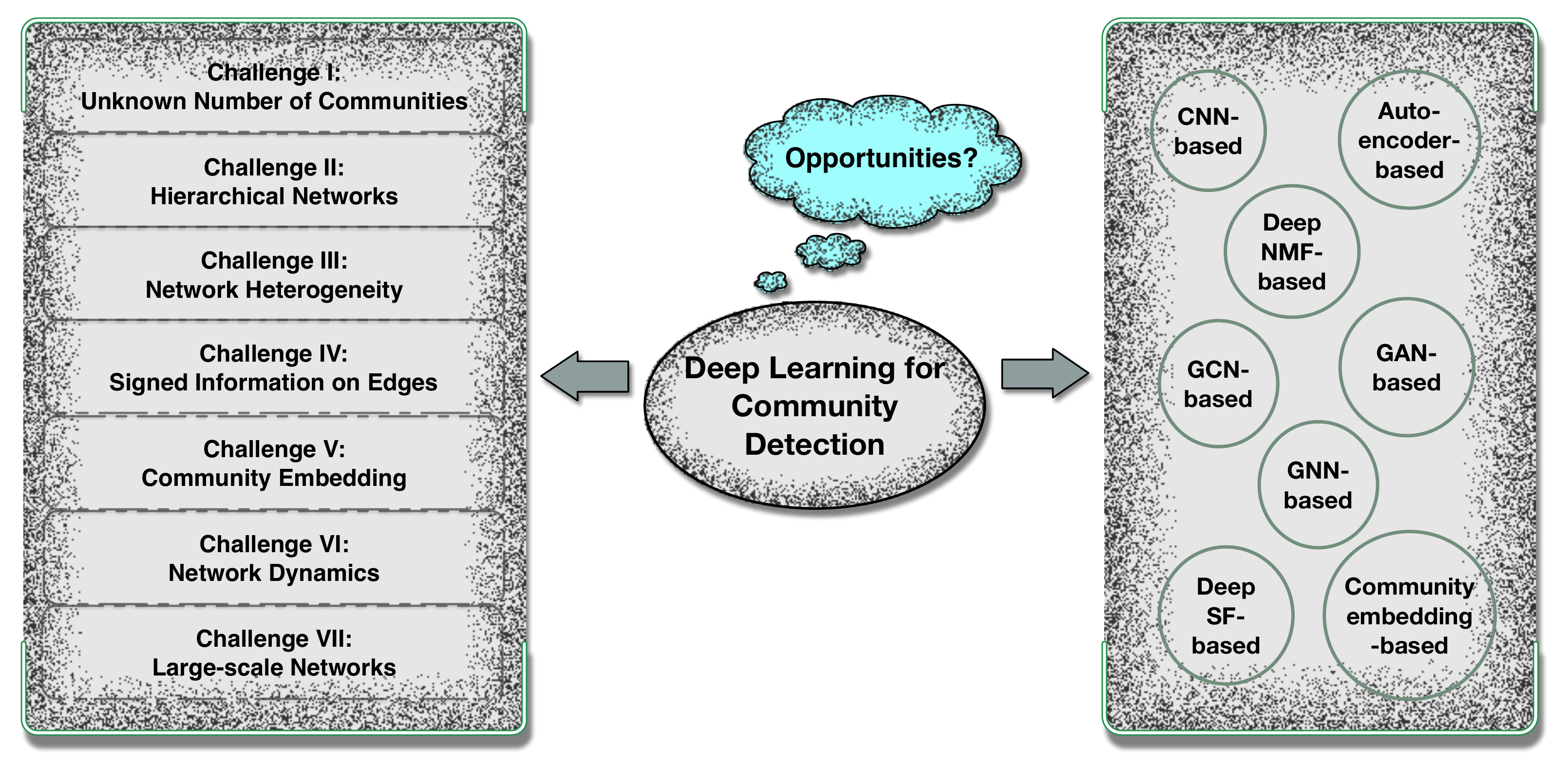}
    \caption{Deep learning for community detection: progress, challenges, and opportunities. This overview is based on concepts from the community detection literature on convolutional neural networks (CNNs), auto-encoders, generative adversarial networks (GANs), deep non-negative matrix factorization (NMF), deep sparse filtering (SF), community embedding, graph neural networks (GNNs), and graph convolutional networks (GCNs). Each approach is discussed in more detail in Section \ref{Sec4}. Each challenge and related opportunity is discussed in more detail in Section \ref{Sec5}.}
    \label{categorization}
\end{figure*}

\section{Why Detection by Deep Learning?}\label{Sec3}
The clear advantage of deep learning for community detection over other machine learning methods is its ability to encode feature representations of high-dimensional data \cite{ZHANG201829}. With graph-structured data, that translates to leveraging the network’s topological information \cite{fortunato2010community}. Deep learning models can also learn the pattern of nodes, neighborhoods, and subgraphs \cite{wu2019comprehensive}. Plus, they are much more resilient to the sparsity associated with large-scale graph data. Lastly, in many real-world scenarios, considering that the majority of nodes are unlabeled and there is little to no prior knowledge of the communities within the data, deep learning is the superior choice for unsupervised learning tasks \cite{tian2014learning}. 

Beyond simply examining network topologies for detecting communities, some strategies also explore semantic descriptions as node features in the data. Traditional community detection approaches are mainly based on an adjacency matrix and a node attribute matrix \cite{Yang6729613,he2017joint}. Deep learning, however, can create much more powerful representations of node attributes and community structures. To this end, recent studies have indicated promising new directions in community detection with deep learning  – for example, the deep learning models in \cite{XIE201850,BHATIA201916,NIPS2019_8342} and modified deep learning models based on community properties \cite{zhang2018cosine,8403293}. 

These studies provide us a glimpse of what deep learning might bring to the future of community detection: 

\begin{itemize}
\item performance improvements; 
\item the capacity to base detection on more and richer features; 
\item the capacity to base detection on both network topology and node attributes for a more robust and better-performing model; and 
\item the ability to detect more complex structures in larger-scale networks. 
\end{itemize}

\section{Community Detection with Deep Learning}
\label{Sec4}

This section provides a technical overview of the recent research progress in deep learning for community detection. Each subsection covers one of the three broad categories of approaches, being deep neural networks, deep graph embedding, and graph neural networks. Within each subsection, we summarize the most influential methods and the contributions of the various frameworks, models, and algorithms. An overview of these methods and their corresponding challenges is provided in Figure \ref{categorization}.

\subsection{Deep Neural Network-based Community Detection}\label{Sec4.1}
Deep neural networks have a natural strength for modeling and capturing comprehensive relationships. The three most popular deep neural network models in the community detection domain are convolutional neural networks (CNNs), auto-encoders, and generative adversarial networks (GANs). Each is summarized in turn below.

\subsubsection{CNN-based Approaches}
The two key components of CNNs are the convolution operation and the pooling operation over the convolutional layers. The convolution operation uses convolution kernels to reduce computation costs. The pooling operation is subsequently applied to the feature mapping to ensure the robustness of CNNs.

Leveraging advancements in CNNs, \cite{Xin2017342} designed a novel CNN model to detect communities in topologically incomplete networks, where some edges are missing when observed from real-world networks. \cite{CNN_SAC} included sparse matrix convolution within a CNN framework specifically to deal with the highly sparse representations associated with adjacency matrices.

\subsubsection{Auto-encoder-based Approaches}
Stacked auto-encoders are a very powerful form of deep neural network model for community detection \cite{yang2016modularity}. 

The inspiration to apply auto-encoders to community detection came from the discovery that auto-encoders and spectral clustering have similar frameworks in terms of a low-dimensional approximation of the spectral matrix \cite{CAO201871}. Focusing on network topology, the approach devised by \cite{Bhatia2018} learns the node clusters via a random walk-based personalized PageRank and fine-tunes the detection by optimizing the modularity of the community structure. To utilize the node attribute information, \cite{CAO201871} proposed a stacked auto-encoder that combines community detection via both the network topology and the attributes of the nodes to enhance the generalization ability of the hidden layer of the deep neural network. To further address the matches between the network topology and node attributes, \cite{Cao_KSEM2018} developed a graph regularized auto-encoder approach by introducing an adaptive parameter as a trade-off control for the matches.

To avoid the need to preset the number of communities, a layer-wise stacked auto-encoder can effectively find centers of communities based on the network structure \cite{Bhatia2018}. Further, this automatic selection mechanism ensures the model assigns nodes strictly based on community criteria. Since this discovery, more and more approaches adaptively learn community structures instead of pre-defining a number. For example, \cite{8594831} introduced a Mixture of Gaussian to capture higher-order patterns from community structures and model the generative process of the observed network to detect communities.

For networks where connections are with positive and negative signs, they are named as signed networks \cite{3330850}. To handle the signed information on edges, a semi-supervised stacked auto-encoder reconstructs the adjacency matrix to represent the signed network so as to the further learning of deep network embedding \cite{8486671}. 

\subsubsection{GAN-based Approaches}
Generative adversarial networks (GANs) involve two deep neural networks competing with each other, which results in fast-adjusting training precision. These methods typically run unsupervised, generating new data with the same statistics as the training set. GANs have been explored to graph representation tasks with great effectiveness \cite{8941296}. 

\cite{Jia_WWW2019} argued that traditional graph clustering-based community detection methods cannot handle the dense overlapping of communities, where one node may belong to more than one community. To this end, they proposed a novel model CommunityGAN that jointly solves overlapping community detection and graph representation learning based on GANs. More importantly, compared with the general graph node representations that have no specific meanings, CommunityGAN has the capability to represent the membership strength of nodes to communities. 

\subsection{Deep Graph Embedding-based Community Detection}
Deep graph embedding is a technique that maps nodes in the network to a low-dimensional vector space, while saving as much structural information as possible in the representations \cite{XUE2019135}. Graph embedding approaches are particularly suited to machine learning tasks based on network analysis, \textit{e.g.,} link prediction, node classification, and node clustering, because they can reference latent features in the representation instead of searching the network. After that, the clustering methods, such as $k$-means, can support the community detection.

\subsubsection{Deep NMF-based Approaches}
The non-negative matrix factorization (NMF) is a group of computational algorithms that factorize a matrix into two matrices with the property that all three matrices have no negative elements. For community detection, NMF approaches approximate the adjacency matrix of a network into the product of two factorized matrices by minimizing the error function for the further clustering task.

\cite{Ye_CIKM2018} proposed a novel deep NMF model that boosts performance by mapping the community structures back to the original network structure during deep learning. By adding node attributes and forming an attributed graph, deep NMF-based community detection only requires an adjacency matrix and node attribute matrix for factorization. In addition, \cite{Li2018community} used NMF for attributed community detection based on node attributes and community embedding in line with the deep feature learning and deep network embedding.

\subsubsection{Deep SF-based Approaches}
Embeddings can encode the input of pairwise relations to avoid searching a sparse adjacency matrix, and hence sparse filtering (SF) is an effective deep feature learning algorithm that only requires one hyperparameter to process high-dimensional inputs \cite{NIPS2011_4334}.  \cite{XIE201850} proposed an efficient network representation approach for network community discovery by way of deep sparse filtering that works particularly well with large-scale networks. An unsupervised deep learning algorithm extracts the network features, which are then used to partition the network.

\subsubsection{Community Embedding-based Approaches}
Traditionally, graph embedding focuses on individual nodes, but communities in networks reflect high-order proximities, such as similar opinions and behaviors. This is an important yet largely under-explored area of graph embedding focused on embedding communities, where the goal is to learn the node distributions of communities in a low-dimensional space. \cite{Cavallari2017} investigated this idea, arguing that this novel and non-trivial strategy could be beneficial to community detection. For example, through a transitional graph embedding method, node distributions could be used to preserve the network structure so as to improve community detection in reverse. \cite{zhang2018cosine} proposed a community-preserving network embedding method to learn network representations. Their performance on community detection demonstrates the superiority. In addition, \cite{8403293} proposed a novel graph embedding model that learns the embeddings of both the nodes and the communities, and its optimization process alternates between community assignment and node embedding, instead of simultaneously solving both tasks.

\subsection{Graph Neural Network-based Community Detection}
GNNs are a technical fusion of graph mining and deep learning, and their recent rapid development is an indication of their power to model and capture the complex relationships in graph-based data. For example, GNN for supervised community detection in \cite{chen2018supervised} introduced a non-backtracking operator to define the edge adjacency. Not only is this approach a feasible way to improve learning performance, but the operator selection is also convenient.

Graph Convolutional Networks (GCNs) inherit the fast learning of CNNs and further improve on that benefit by integrating a probabilistic model that considers the probability distribution of the entities in the network. For example, \cite{jin2019graph} combined a Markov random field with an attributed network of semantic information to support semi-supervised learning, while \cite{oleks2019} integrated a Bernoulli–Poisson probabilistic model with a GCN for overlapping community detection such that the convolutional layers can recognize complex network patterns.

\section{Challenges and Opportunities}\label{Sec5}
Although we have witnessed the rapid developments in deep learning for community detection in recent years, especially the last five, there are still issues that need better solutions and challenges that remain unresolved. As with most gaps in our knowledge, these problems provide opportunities for future study. In this section, we discuss seven broad challenges facing the community, beginning with, arguably, the longest-standing issue in community detection.

\subsection{Challenge I: An Unknown Number of Communities}
The challenge caused by the need to know the number of communities in advance has existed for a very long time, and, despite some professed solutions, this problem still has not been fully resolved. In machine learning, community detection is usually approached as an unsupervised clustering task because most of the data extracted from real-world networks do not have labels. This leads to a catch-22: the data needs to be labeled to determine the number of communities, but the number of communities needs to be known before the data can be labeled. Deep learning methods work around this problem to some extent by clustering nodes according to similarity in one or more latent spaces. However, the clustering algorithms still need to know the final number of clusters in advance.

\textit{Opportunities}: 
To this point, an effective solution is to calculate the number of communities by analyzing the network topology, such as \cite{Bhatia2018,BHATIA201916} based on random walk-based personalized PageRank. However, this type of method cannot guarantee that every node in the network is assigned to a community. Thus, a full and complete solution to this problem is yet to be found.

\subsection{Challenge II: Hierarchical Networks}
Hierarchical networks are made up of layered networks, where each layer of networks shares specified functions. Hence, the community detection strategy must be able to extract layer-wise representations. The challenges involved in this include distinguishing different relationship types, such as horizontal and vertical, and managing varying levels of sparsity in different layers.

\textit{Opportunities}:
\cite{song2018improved} proposed a multi-layer DeepWalk with creating inter-layer edges to exploit the dependencies across different layers while preserving hierarchical structures. It learns representations for each node in every layer which are fine-tuned by a refinement strategy. Another possible solution is to simultaneously optimize for common representations applicable to all layers and local representations preserving the layer-specific network structure. Moreover, the scalability of the scheme with respect to the number of layers in the hierarchy is questionable and should be considered in devising novel solutions. In addition, novel models are desired to distinguish different kinds of connections in hierarchical networks. Thus, there is much work to be done before we have a deep learning method for detecting hierarchical network communities.

\subsection{Challenge III: Network Heterogeneity}
Network heterogeneity refers to networks that contain significantly different types of entities and relationships, which means the strategies used for homogeneous networks do not necessarily work. In particular, the different probability distributions associated with each type of entity need to be addressed in the design of models and algorithms.

\textit{Opportunities}: 
To date, few deep learning approaches consider network heterogeneity. Of the work we surveyed,  \cite{Chang2015} tackled this issue with a nonlinear embedding function to capture the complex interactions between heterogeneous network entities. However, their method neglects the different semantics of relationships between nodes \cite{Fu_CIKM17}. The future opportunities for community detection in heterogeneous networks may include: 1) deep graph embedding models and supporting algorithms; 2) specific deep learning models with novel training processes to learn heterogeneous graph properties within the hidden layers; and 3) novel models that can exploit different types of relationships among nodes.

\subsection{Challenge IV: Signed Information on Edges}
Many real-world networks have signed edges representing positive or negative relationships. The challenge is that these relationships need to be treated in different ways.

\textit{Opportunities:}
One possible solution is to incorporate positive and negative edges by designing a random walk procedure. \cite{Hu2019} followed this idea and developed a sparse graph embedding model based on word embedding. But their method does not perform as well as the baseline spectral methods for some small real-world signed networks. Another possible solution is to reconstruct the adjacency matrix representation of a signed network. However, this raises other problems because, in the real-world, adjacency connections are overwhelmingly positive. \cite{8486671} imposed a penalty to ensure their stacked auto-encoder model focuses more on reconstructing the rarer negative edges over the abundant positive edges. However, without prior knowledge of most of the relationships in a community, which is realistic in many situations, this approach will not work. Thus, efficient methods for unsupervised community detection in signed networks are still required.

\subsection{Challenge V: Community Embedding}
Community embedding looks to create representations of communities instead of representations for each individual node \cite{Cavallari2017}. Thus, focus shifts to community-aware high-order proximity instead of the 1- or 2-order proximity associated with node neighborhoods. This is an emerging research area with three main challenges to overcome: 1) high computational costs; 2) relational evaluations between node and community structures; and 3) other problems when applying deep learning models, such as distribution shifts across communities.

\textit{Opportunities:}
From our survey, we intuited an intelligent way to support community embedding by automatically selecting the representation modules on nodes and/or communities. To this end, we suggest the following research objectives: 1) explore how to integrate community embedding into a deep learning model; 2) determine how to directly embed community structures for a range of benefits, such as fast computation; and 3) devise a method for optimizing the hyperparameters in an integrated deep community detection learning model.

\subsection{Challenge VI: Network Dynamics}
Changing dynamics can affect either the network topology or the node attributes, each of which must be handled in their own way. Topological changes, such as adding or deleting a node or edge, not only cause changes in a local community, they can also lead to global changes across an entire network \cite{liu2019detecting}. With dynamic networks, deep learning models need to be re-trained over a series of snapshots. The technical challenge with the temporal attributes of a dynamic network lies in the deep feature extraction of dynamics.

\textit{Opportunities:}
Deep learning methods have not yet been well developed for detecting communities with dynamic spatial and temporal properties. Therefore, future directions of research include: 1) detecting and recognizing the spatial changes over communities; 2) learning deep patterns that embed both temporal feature and community structure information; and 3) developing a unified deep learning approach for community detection that can simultaneously handle both spatial and temporal features.

\subsection{Challenge VII: Large-scale Networks}
Today, large-scale networks can contain millions of nodes, edges, and structural patterns and can also be highly dynamic, as networks like Facebook and Twitter demonstrate. This presents many areas of development that need to be addressed before truly large-scale community detection can become a reality. For instance, large-scale networks may have their inherent scale characteristics, such as scale-free (\textit{i.e.,} a power-law degree distribution) in social networks. Such particular distribution can influence the performance of deep learning in community detection. Scalability seems to be another key issue to enable deep learning to detect communities in large-scale networked environments. Constantly changing network typologies further increases the detection difficulty. Overall, deep community detection in large-scale networks involves all six of the preceding challenges as well as the challenge of scalable learning.

\textit{Opportunities:} 
To fully use the rich information in large-scale networks, novel unsupervised clustering algorithms are needed that have lower computational complexity and greater flexibility. Distributed computing architectures are popular in large-scale machine learning community. Thus, one possible direction is to develop a robust deep learning detection approach that can achieve an high-performance collaborative computing. On the other hand, regarding the high-dimensional adjacency matrix, the key strategy of dimension reduction commonly used in deep learning (\textit{i.e.,} matrix low-rank approximation) does not work with large-scale networks, and even the current distributed computing solutions are still too expensive. Therefore, novel deep learning frameworks, models, and algorithms are in high demand. As probably the biggest challenge in community detection, those frameworks need to far exceed the current benchmarks in precision and speed.

\section{Conclusions}
\label{Sec6}
Nowadays, all of us live, work, and play in an intricate system of many communities. Likewise, as our knowledge of the world grows, we find objects perform more than one function, ideas serve more than one purpose, and there is far greater interconnectedness between all things than we previously imagined. Detecting these communities and their inherent functions and features helps us to comprehensively understand our surrounding environment and discover complex relationships to describe and explain social phenomena. 

Traditional community detection methods have typically relied on statistical inference and conventional machine learning methods, such as stochastic block model and spectral clustering. But advancements in deep learning have given rise to community detection strategies with more powerful capacity of handling high-dimensional graph data with impressive performance. 

In this survey, we reviewed the technical trends and current state-of-play in model and algorithm development in three broad deep-learning approaches to detecting communities in various scenarios. We also identified seven vital challenges that need to be overcome for community detection through deep learning to fully mature. 

To some extent, these challenges serve as guidelines for the next generation of community detection. Although many researchers are actively engaged in addressing these problems, we point to the numerous opportunities that still exist to contribute to the development of this important ever-evolving field.

\section*{Acknowledgments}
This work is supported in part by ARC DECRA under grant DE200100964, in part by the 2018 Collaborative Project between MQ and Data61, in part by the MQEPS under grant 96804590, in part by the NSFC under grants 61872360 and 61976162, and the Youth Innovation Promotion Association CAS under grant 2017210, and in part by NSF under grants III-1526499, III-1763325, III-1909323, and CNS-1930941. F. Liu and S. Xue contributed equally to this work.

\clearpage
\setlength{\baselineskip}{0.5pt}
\small
\bibliographystyle{named}
\bibliography{ijcai20}

\end{document}